\documentclass[twocolumn, 10pt]{article}
\usepackage{amsmath,amssymb}
\usepackage{graphicx}
\usepackage{float}
\usepackage{diagbox}
\usepackage[format=hang,labelfont=bf,textfont=small,singlelinecheck=false,justification=raggedright,margin={12pt,12pt},figurename=Fig.]{caption}
\usepackage{titlesec}
\usepackage{textcomp}
\usepackage{cite}
\usepackage{float}
\usepackage{makecell}
\usepackage{hyperref}
\usepackage{subcaption} 
\usepackage[export]{adjustbox}
\usepackage{etoolbox}
\makeatletter
\def\affiliation#1{\gdef\@affiliation{#1}}
\def\abstract#1{\gdef\@abstract{#1}}
\def\graphabst#1{\gdef\@graphabst{#1}}
\def\keywords#1{\gdef\@keywords{#1}}
\def\corresp#1{\gdef\@corresp{#1}}

\newcommand{\MakeTitle}{
  \newpage
  \null
  \vskip 2em%
  \begin{center}%
  \Large \@title\par
  \vskip 1em%
  \large \@author
  \end{center}
  \noindent\@affiliation\par
  \vskip 1em%
  \noindent\@corresp\par
  \vskip 1em%
  \noindent\@abstract\par
  \vskip 1em%
  \noindent\@keywords\par
}
\makeatother

\makeatletter
\patchcmd{\@maketitle}{\raggedright}{\centering}{}{}
\patchcmd{\@maketitle}{\raggedright}{\centering}{}{}
\makeatother

\newcommand*{\TitleFont}{%
      \usefont{\encodingdefault}{\rmdefault}{}{n}%
      \fontsize{18}{12}%
      \selectfont}

\titleformat{\section}
  {\normalfont\fontsize{10}{11}\bfseries}{\thesection.}{2pt}{}
  \titlespacing*{\section}{0pt}{12pt}{6pt}

\titleformat{\subsection}
  {\normalfont\fontsize{10}{10}\bfseries}{\thesubsection.}{2pt}{}
  \titlespacing*{\subsection}{0pt}{6pt}{0pt}
  
\titleformat{\subsubsection}
  {\normalfont\fontsize{10}{10}\bfseries}{\thesubsubsection.}{2pt}{}
  \titlespacing*{\subsubsection}{0pt}{6pt}{0pt}
  
\normalsize

\title{\TitleFont Can magnetic field be used to reduce cosmic charged particles background to low energy particle detectors?}

\author{Kolahal Bhattacharya}

\affiliation{
\begin{center}
Manipal Centre for Natural Sciences\\ 
Centre of Excellence\\ 
Manipal Academy of Higher Education, Manipal 576104. India
\end{center}
}

\corresp{$^{\ast }$kolahal.bhattacharya@manipal.edu 
}

\abstract{\textbf{Abstract}: The possibility to reduce the background due to cosmic ray charged particles by the use of magnetic field in the ground based low energy particle detectors is explored. The degree of reduction of cosmic rays as a function of the magnetic field strength and its depth is quantified.}


\begin{document}

\onecolumn
\MakeTitle
\section{Introduction}
Cosmic rays pose a serious background to a vast majority of the ground-based nuclear and particle physics experiments. For example, many current generation neutrino physics experiments attempt to observe muons and electrons generated in the neutrino events to resolve the unknown questions in neutrino physics, e.g. CP violation, neutrino mass hierarchy, or existence of sterile neutrinos etc. Since a major component of the cosmic rays are muons, removal of the muon signature due to cosmic background is an absolute necessity for these experiments. For this reason, it is common to design experimental facilities in underground (e.g. DEAP-3600 in SNOlab~\cite{boulay2012deap}, or ICECUBE in Antarctica~\cite{abbasi2012design}), or to use the plastic scintillator-based veto method to effectively reduce the background (e.g. MicroBooNE~\cite{acciarri2017design} and mu2e~\cite{woodward2019fabrication} in Fermilab). Many upcoming experiments (e.g. INO~\cite{kumar2017invited}, and far detector of DUNE~\cite{abi2018dune}) will also place their detectors deep underground to this end. However, building the underground laboratories from scratch is expensive. Therefore, many experiments in past (and present) were (are) conducted where mine was (is) already in existence, e.g. the `proton decay' experiment in Kolar gold field (in India), or the SNOlab in Sudbury, Canada. On the other hand, the paddle scintillator-based veto method can be used effectively in conjunction with the beam pulse window to reduce the cosmic background. This
method can be very successful in accelerator-based neutrino experiments where there is definite time window of events to occur. However, for a ground based non-accelerator experiment, it is not a very practical method where timings of the events are not known.

In this context, the question whether it is possible to reduce a significant fraction of the cosmic ray background by deflecting them using magnetic field is addressed in this paper. This looks appealing, because this method may be able to reduce the charged particle component of the cosmic rays for both accelerator-based as well as non-accelerator based experiments. At the same time, it appears that this may require extremely high magnetic field to have any practical utility in any particle detectors. No quantified results were found in the scientific literature to address this question. Even if the method is not practical, due to the requirement of very strong magnetic field over a large volume, that needs to be addressed. In this paper, the strength and volume of the field needed to achieve a given degree of reduction will be estimated. Also, the threshold of the momenta of the cosmic ray charged particles which cannot be deflected away will be investigated. These details can be used to decide whether magnetic field can be employed in specific future experiments. If it is possible, then one can try to figure out the way to accomplish that. For example, it may be possible to achieve the experiment's desired degree of reduction of cosmic ray charged particles by building a very shallow (50 m deep) underground detector and a magnet system constructed above the ground. The paper will also discuss the currently available technologies for generating necessary magnetic field and other indirect benefits of the reduction of cosmic muons. To address these points, a thought experiment, described in the next section~\ref{Sec2}, has been performed using GEANT4~\cite {agostinelli2003geant4} simulation program. The outcomes of this experiment, will hopefully throw some light on the topic.


\section{Set up for thought experiment}~\label{Sec2}
The basic configuration of the thought experiment is shown in the following Figure~\ref{fg1a}. The magnet system to deflect the charge particle component of the cosmic rays must be placed over the particle detector. This may be able to reduce the background to the detector placed underneath. However, the cosmic muons can also reach the detector from sides. Therefore, it may be helpful to use a magnet that spans a large area and to place it on top or over the particle detector, shown as a brown right rectangular parallelepiped in Figure ~\ref{fg1b}. 

\begin{figure}[h]
\centering
\captionsetup{justification=centering}
\begin{subfigure}{.15\textwidth}
  \centering
  \includegraphics[height=4.0 cm, width= 5.0 cm]{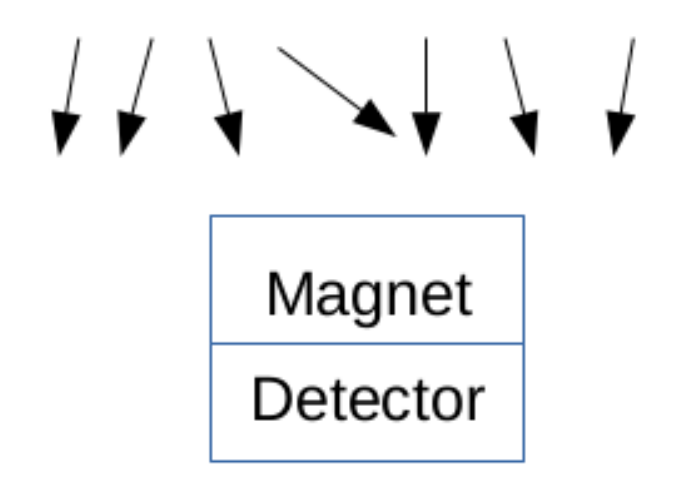}
  \caption{}
  \label{fg1a}
\end{subfigure}
\begin{subfigure}{.8\textwidth}
  \centering
  \captionsetup{justification=centering}
  \includegraphics[height=3.5 cm, width= 7.5 cm]{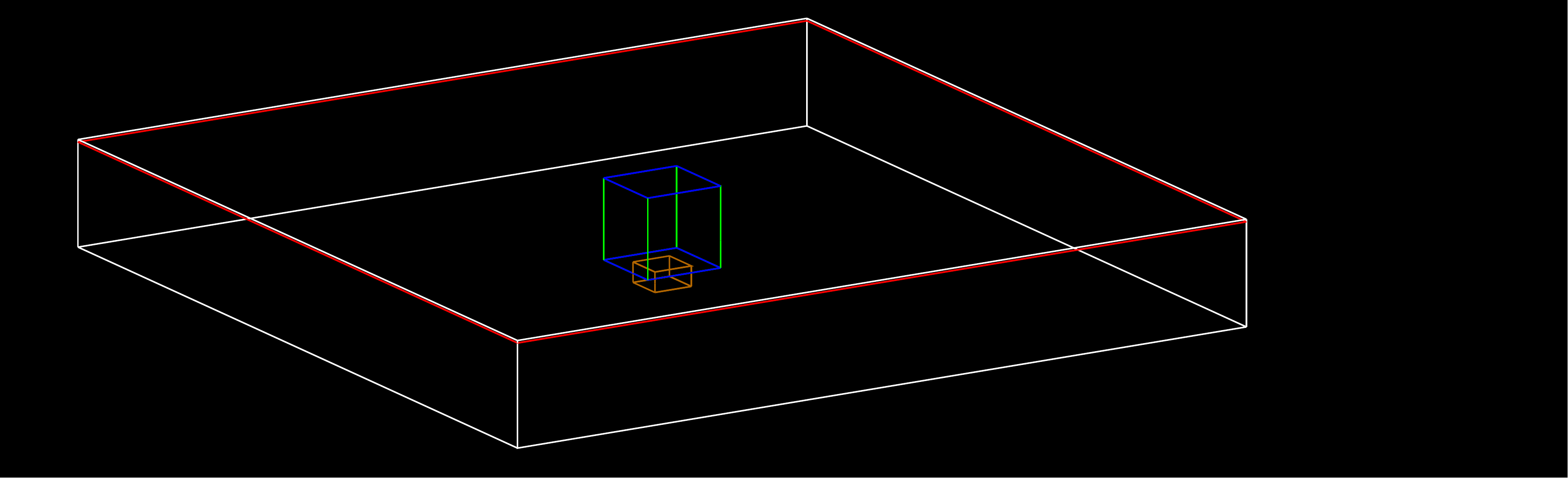}
  \caption{}
  \label{fg1b}
\end{subfigure}
\caption{(a) Magnet system as a buffer between the down headed cosmic rays and the particle detector, (b) the system is shown as a green box lined with blue boundaries at the top and bottom surfaces. It is placed on top of the particle detector (brown) placed underneath. The whole setup is placed below a large reference plane with red boundary (upper plane of the magnet coincides with the reference plane).}
\label{fig:1}
\end{figure}
To exclusively study the effect of magnetic field, the field is assumed to operate in vacuum, so that energy loss and generation of secondary particles do not occur when cosmic rays traverse through the magnet. Practically, the magnet system can be composed of iron, or neodymium magnets or the superconducting magnets and must be assembled within a box made of shielding material (e.g. Mu-metal, iron) to ensure that the magnetic field lines do not penetrate the detector placed underneath. The field lines should return from the far north pole to the far south pole through the wall of this shielding box.

A typical ground based particle detector can be placed in an underground pit, and the magnet system may be placed on top of it, supporting on the surrounding ground level. The field should point in the horizontal direction, because a component parallel to the vertical direction is not helpful to deflect the muons coming vertically down.
We have used the CRY~\cite{hagmann2007cosmic} cosmic ray generator to simulate the cosmic ray shower (muons $\mu^\pm$, electrons $e^\pm$, pions $\pi^\pm$, kaons and protons) on top of the reference plane of Figure~\ref{fg1b}, lined with red boundary. This results in contribution of cosmic rays from the sides as well, in addition to the usual vertical flux. The dimension of this reference plane is taken as 100 m$ \times$100 m and the dimension of particle detector is taken as $\Delta x=$5 m, $\Delta y=$5 m and $\Delta z=$ 2.5 m (vertical direction). The size of the magnet system (shown as the green box in Figure~\ref{fg1b}) is ($\Delta x_{mag}, \Delta y_{mag}, \Delta z_{mag}$). These variables will be varied to find out the residual flux leaked into the particle detector (brown) placed underneath. If $\Delta x_{mag}=\Delta y_{mag}=\Delta z_{mag}=10$ m, then the boundaries of the reference plane subtends a zenith angle of $\tan^{-1}\frac{50}{10+ 2.5}\approx76^o$. If $\Delta z_{mag}$ is less, the zenith angle coverage will be higher. If $\theta$ denotes the zenith angle, then the cosmic muon flux approximately drops as $\cos^2\theta$~\cite {shukla2018energy}. At larger $\theta$, the flux reduces significantly and majority of the cosmic muon flux is contained in lower $\theta$. So, these dimensions should be good enough to represent realistic muon flux leaking into the detector. 

\subsection{Case I: No magnetic field}
Generation of $3$ million events in CRY at ground level (i.e. zero altitude) results in a shower of about $3.7$ million tracks raining down on the 100 m$\times$100 m wide reference plane (additional tracks are generated when cosmic rays collide with the atmosphere and multiple daughter particles are produced) in $\sim$ 2.379 seconds. This shower is passed through the magnet-detector system using the GEANT4 package and the number of tracks entering the detector is counted. If there is no magnetic field, the cosmic rays will simply pass through without any deflection. The number 3 million is selected such that about 10 thousand cosmic muons leak into the detector in the absence of any magnetic field in the magnet system. This is shown in the following table~\ref{tabnomag}:

\begin{table}[H]
    \centering
    \begin{tabular}{ccccc}\hline
    Depth & \# cosmic rays & \# muons & \# electrons & \# protons\\ \hline
    1 m   & 13943 & 10862 & 2903 & 148\\
     2 m   & 13917 & 10849 & 2892 & 142\\ \hline
    \end{tabular}
    \caption{Out of about 3 million charged cosmic ray particles simulated by CRY, about 10 thousand muons enter the particle detector from all directions in about 2.379 seconds. Pions and kaons are almost negligible in number.}
    \label{tabnomag}
\end{table}

In fact, the number of cosmic ray charged particles entering the detector is roughly proportional to the fraction of the effective area of the detector to the aperture of the reference plane, i.e. $\frac {5\times5}{100 \times100}\sim0.0025$. This accounts for the order of magnitude of the number of cosmic ray particles entering the detector. In reality, it is little more than that, due to the influx of the cosmic rays from side walls, as found in the above table~\ref{tabnomag}. The following figure~\ref{2a} shows the end $z$ coordinate of the tracks entering the detector for a magnet system of depth 1 meter.
\begin{figure}[ht]
\centering
\begin{subfigure}{.45\textwidth}
  \centering
  \captionsetup{justification=centering}
  \includegraphics[height=5.0 cm, width= 7.5 cm]{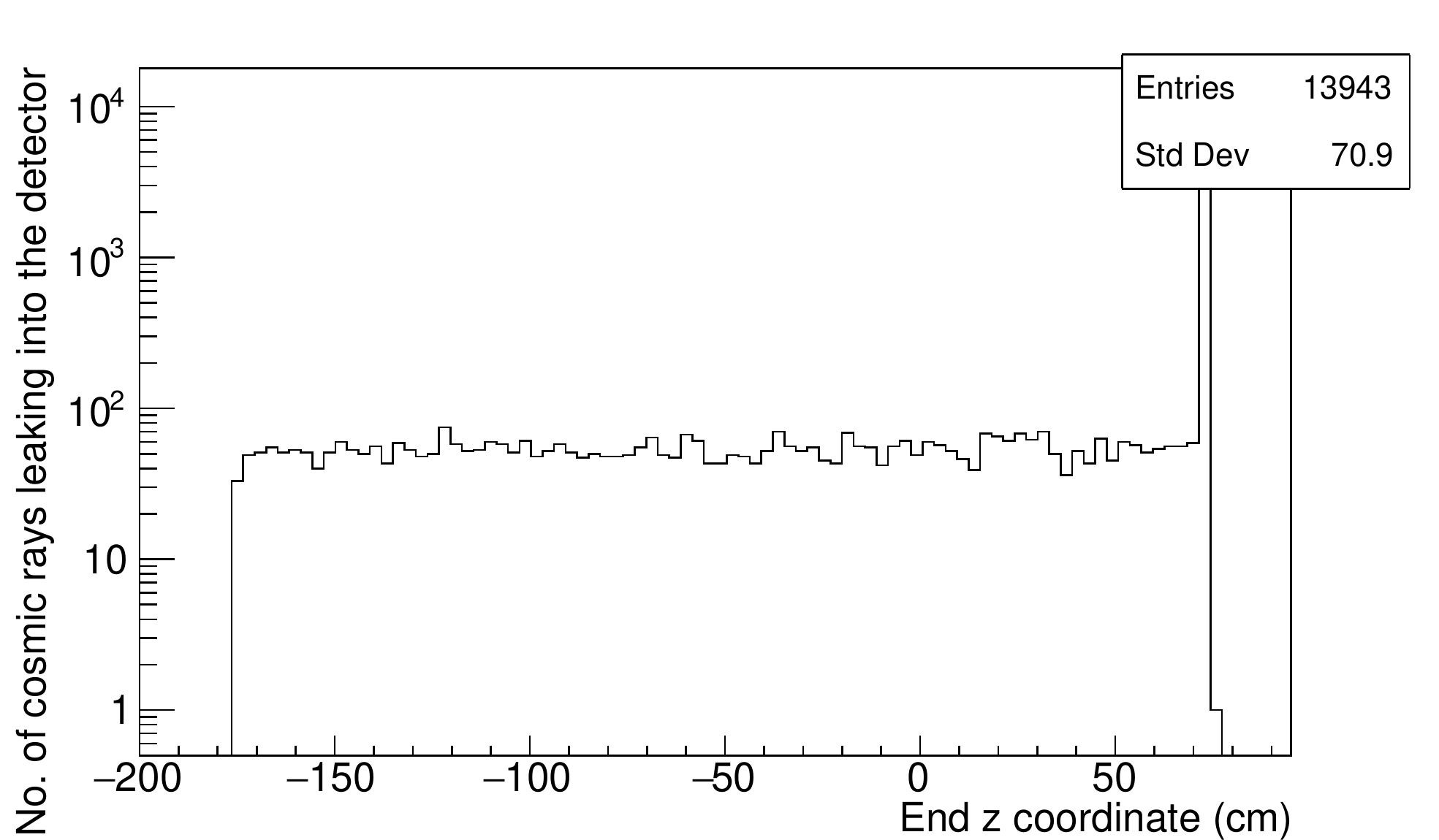}
  \caption{}
  \label{2a}
\end{subfigure}
\hspace{0.0 cm}
\begin{subfigure}{.45\textwidth}
  \centering
  \captionsetup{justification=centering}
  \includegraphics[height=5.0 cm, width= 7.5 cm]{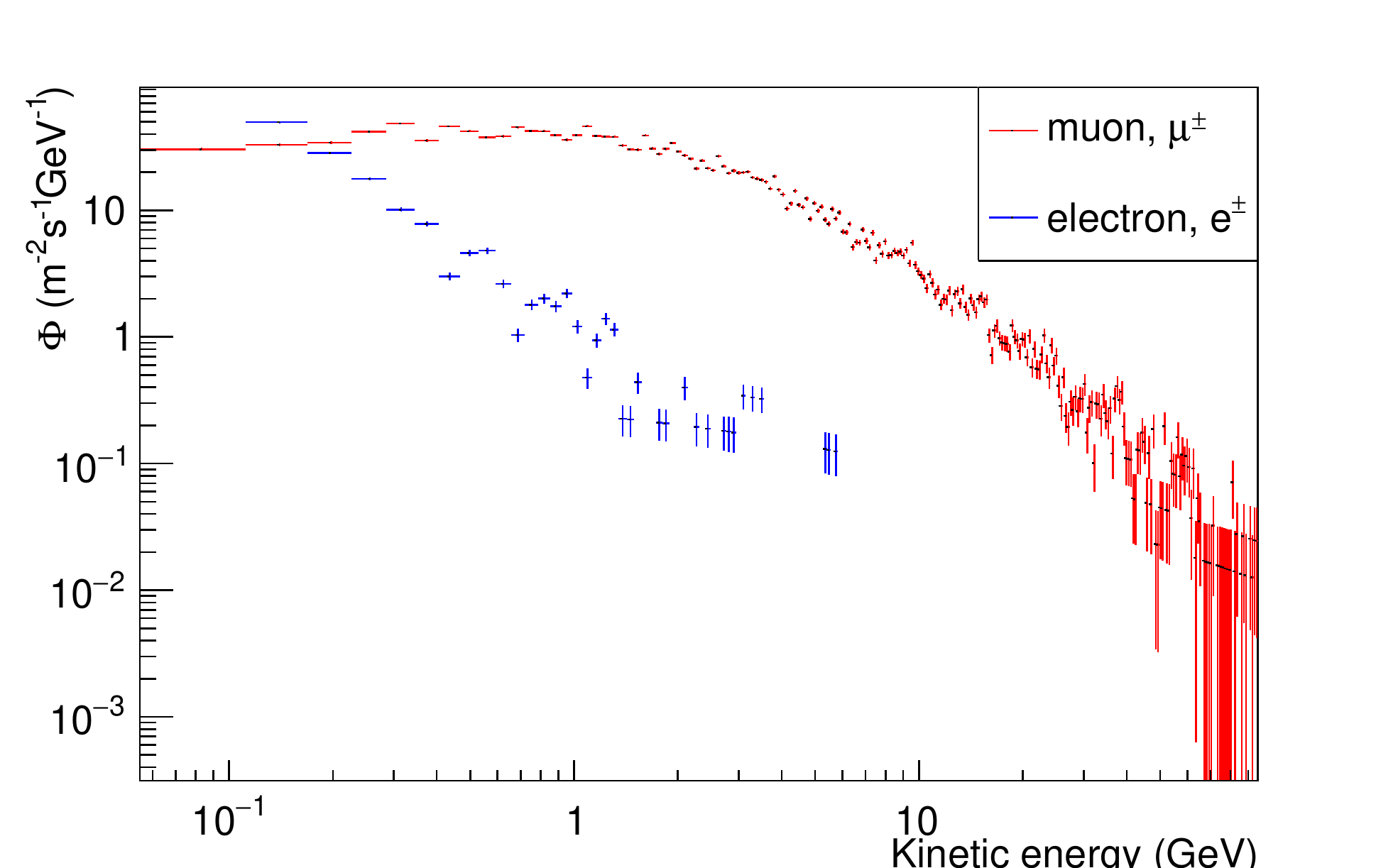}
  \caption{}
  \label{2b}
\end{subfigure}
\caption{(a) Cosmic ray tracks are killed as they enter the particle detector from top or from sides. For 1 meter deep magnet, the height of the magnet and detector system is $\sim$(1+2.5) m $\sim$ 3.5 m. The origin of GEANT4 system is at the midpoint.
The spike in this plot corresponds to the majority of the cosmic rays hitting the detector at its top surface at $\sim$(2.5 m - 1.75 m)$\sim$ 0.75 m. (b) Spectra of residual flux of muons and electrons entering the particle detector.}
\label{fig2:NoMag}
\end{figure}
From table~\ref{tabnomag}, it is seen that the main background comes from muons and electrons. Spectra of these particles at the point of entering the particle detector in absence of magnetic field are shown in the figure~\ref{2b}. 

\subsection{Effect of introducing magnetic field}
The following study assumes uniform magnetic field and without any loss of generality, its direction is taken as the positive $y$ direction. The degee of reduction of the flux with the use of magnetic field will be quantified next. It is clear that there are several parameters in this problem:\\
(a) depth $\Delta z_{mag}$ of the magnet system; a higher depth might allow deflection of a higher fraction of muons.\\
(b) strength of magnetic field. \\
(c) the transverse area of the magnet system; how much more area compared to the aperture of the detector is needed?\\
(d) energy spectrum and composition of the particles (i.e. the relative fraction of different particles, e.g. muon, electrons etc.) leaking into the detector.\\
(e) material of the magnet system and its effect on the above parameters.

All these details will be useful to throw light on different aspects of the problem and may give hints towards future building of experiments. Below these parameters are investigated in details.


GEANT4 simulation was performed for 10 m $\times$ 10 m wide magnet systems of different strengths (1-5 tesla) and depths (1-10 m). The following figure~\ref{3a} shows the energy spectrum of muons, electrons and protons when 1.5 tesla field is applied through a magnet of depth 1 meter. Muons dominate the spectrum, specifically at higher energy. However, in comparison with the no magnetic field case (figure~\ref{2b}), the charged 
particle background is seen to be suppressed at the lower energy bins.
\begin{figure}[H]
\centering
\begin{subfigure}{.45\textwidth}
  \centering
  \captionsetup{justification=centering}
  \includegraphics[height=5.0 cm, width= 7.5 cm]{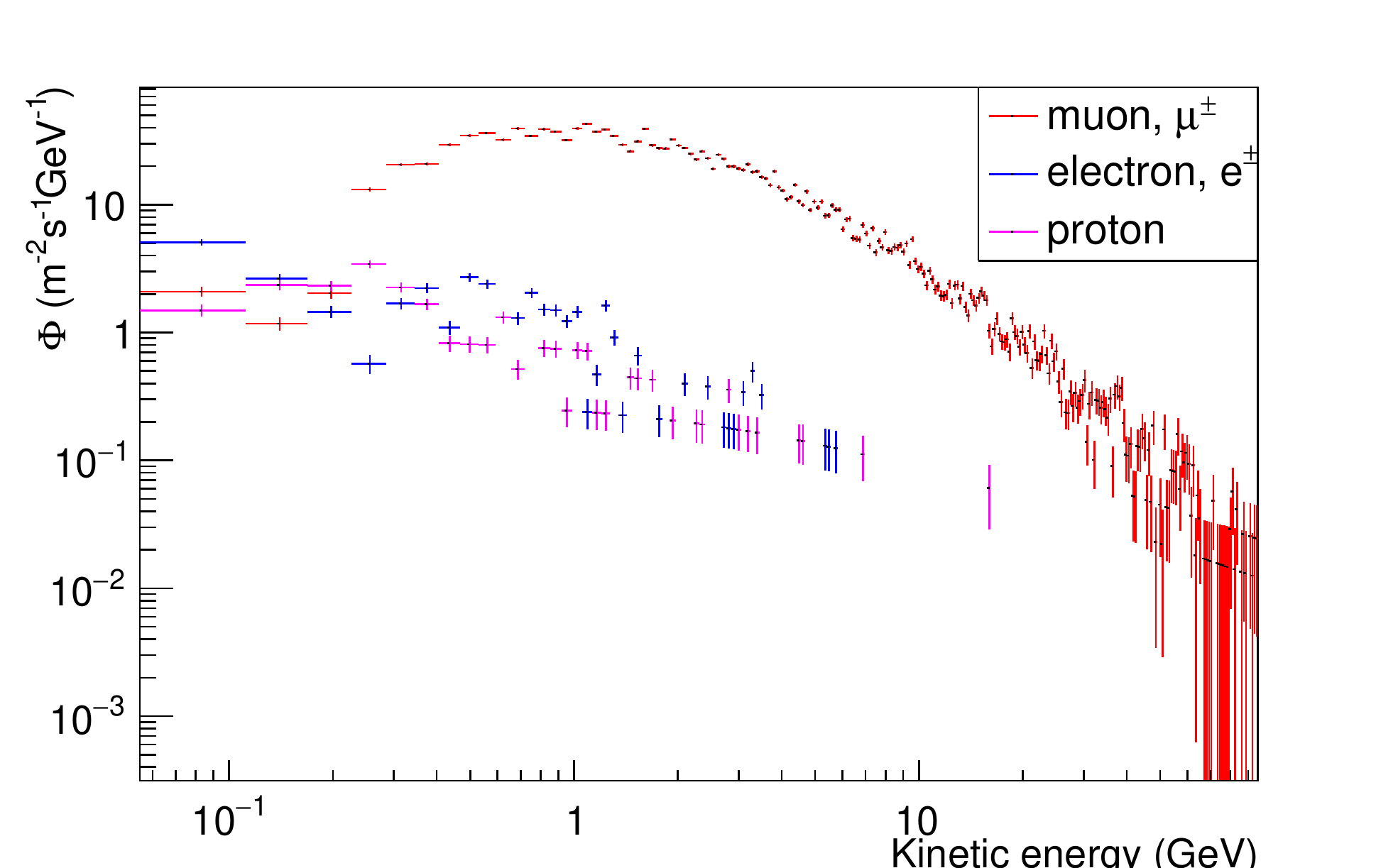}
  \caption{}
  \label{3a}
\end{subfigure}
\hspace{-0.5 cm}
\begin{subfigure}{.45\textwidth}
\centering
\captionsetup{justification=centering}
    \begin{tabular}[b]{cccccc}\hline
Depth (m) & Field (T) & Total & protons & $e^\pm$ & $\mu^\pm$ \\ \hline
1 & 1.5 & 10299 & 99 & 267 & 9908 \\
3 & 1.5 &  7735 & 43 & 212 & 7450 \\
5 & 1.5 &  6381 & 33 & 155 & 6170 \\
7 & 1.5 &  5797 & 19 & 148 & 5609 \\ \hline
1 & 1.0 & 10837 & 119 & 336 & 10354 \\
1 & 2.0 &  9824 & 85 & 239 &  9472 \\
1 & 3.0 &  8877 & 70 & 215 &  8563 \\
1 & 5.0 &  7276 & 40 & 177 &  7030 \\ 
\hline
    \end{tabular}
\caption{}\label{tab2}
\end{subfigure}
\caption{(a) Energy spectra of muons, electrons and protons leaking into the detector for 1.5 tesla field operating at 1 meter depth. (b) Effect of varying depth and magnetic field on the composition of cosmic ray spectrum leaking into the detector.}
\label{fig:2}
\end{figure}
The relative percentage of different particles, i.e. composition of the residual cosmic ray charged particle flux does not change significantly if the strength and/or the depth of the magnet are varied. This is shown in table~\ref{tab2}. However, overall decrease in number is observed with the increase in field strength. The trend is shown for 3 tesla and 5 tesla field operating at 1 m depth in figure~\ref {4a}.
\begin{figure}[H]
\centering
\begin{subfigure}{.45\textwidth}
  \centering
  \captionsetup{justification=centering}
  \includegraphics[height=5.0 cm, width= 7.5 cm]{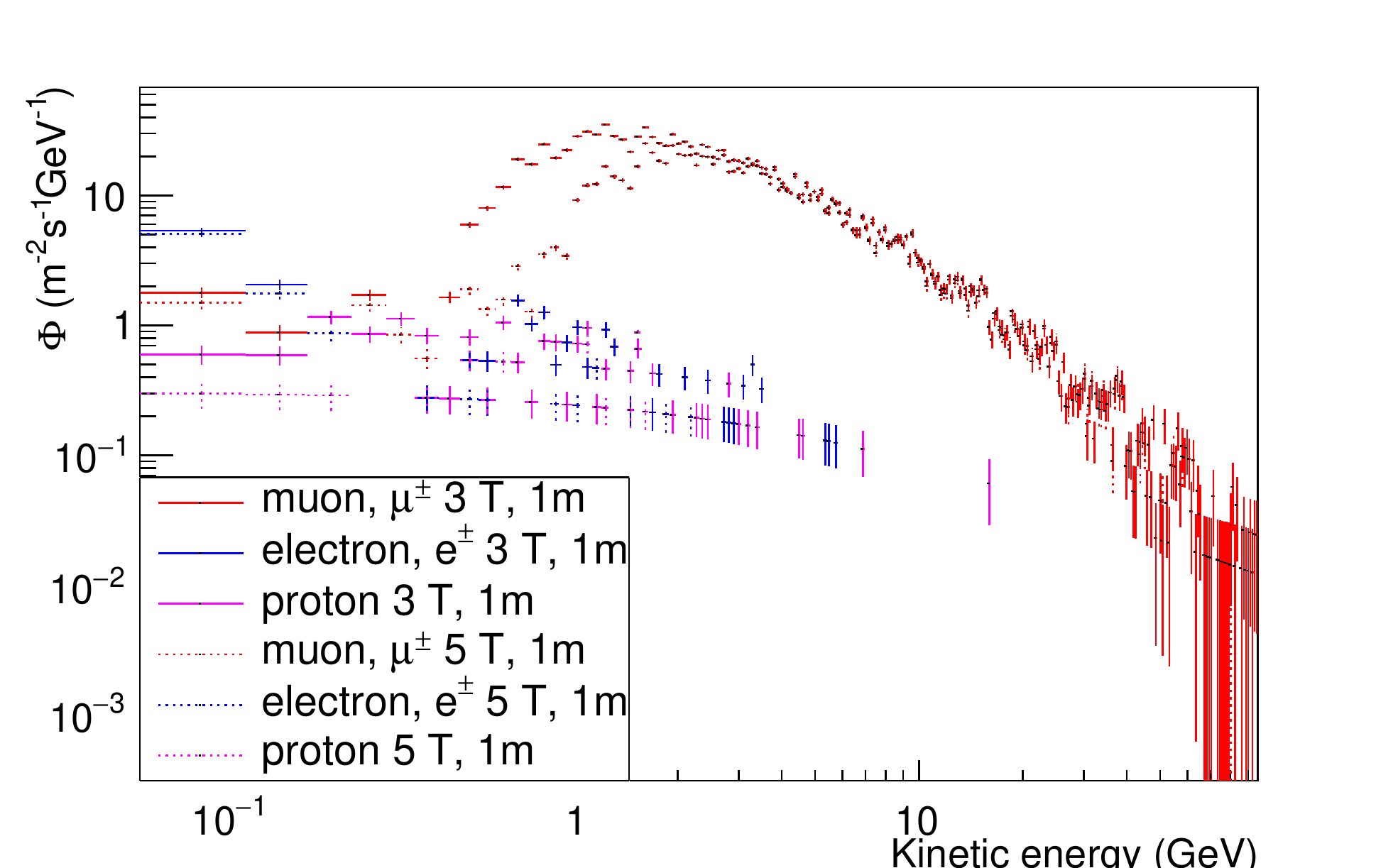}
  \caption{}
  \label{4a}
\end{subfigure}
\hspace{0.0 cm}
\begin{subfigure}{.45\textwidth}
  \centering
  \captionsetup{justification=centering}
  \includegraphics[height=5.0 cm, width= 7.5 cm]{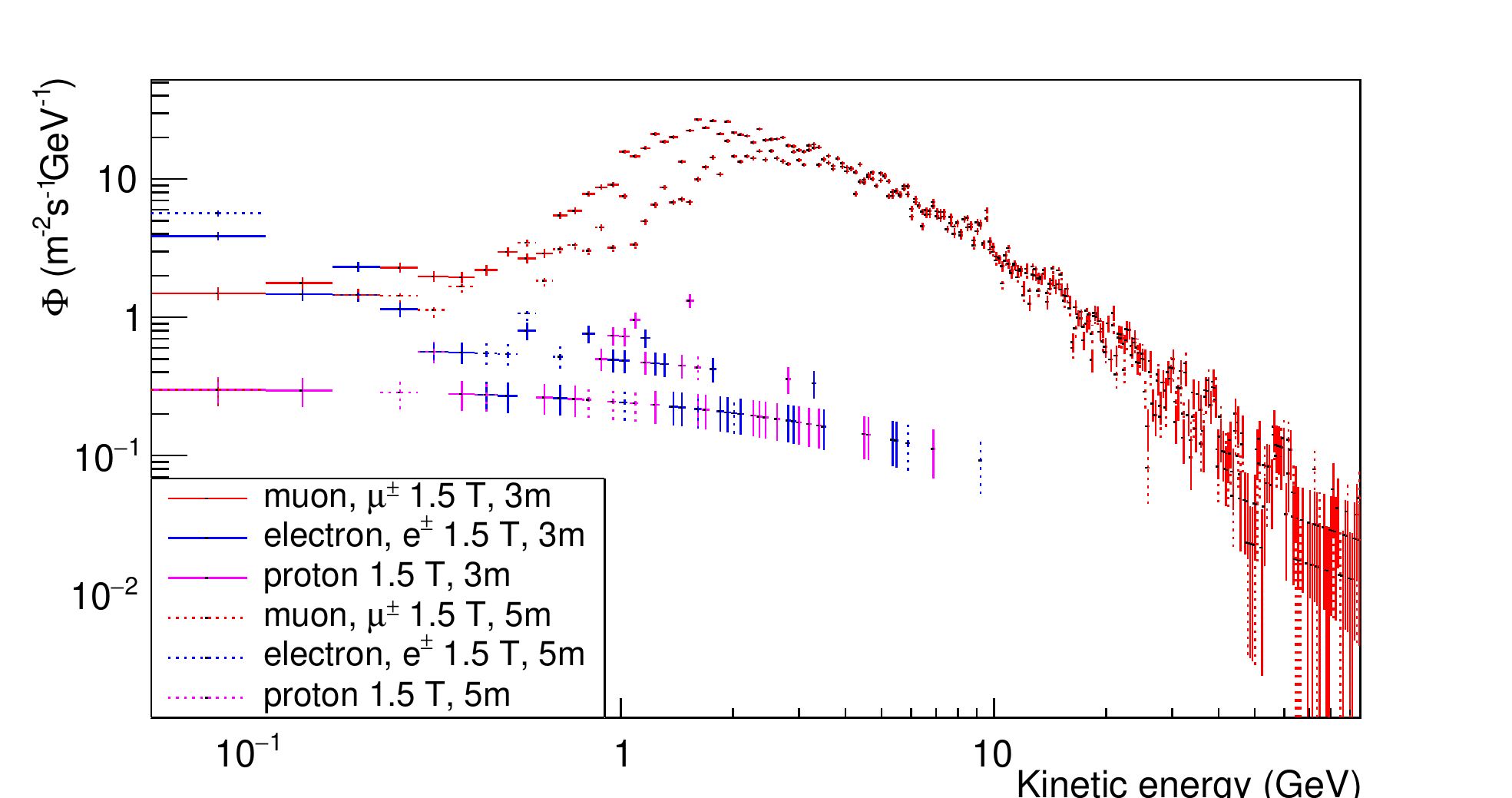}
  \caption{}
  \label{4b}
\end{subfigure}
\caption{Effect of increase in (a) magnetic field and (b) depth on the cosmic ray charged particle spectrum.}
\label{fig4}
\end{figure}
Figure~\ref{4b}, on the other hand, shows the spectra of these particles if depth of the magnet is increased without changing the field strength. The reduction can be understood in the following way: the depth of the magnet system is the effective `clearance distance' the charged particles have before they bend away. Evidently, the low energy muons will be deflected away at a lower depth whereas high energy muons will need higher depth, if both pass through a system of the same magnetic field. 

\subsection{Relative reduction with respect to no magnetic field case}
One way to quantify the degree of reduction of the cosmic ray muons due to magnetic field is to take the ratio of the number of all the cosmic ray muons leaking into the detector in the presence of the  magnetic field to the number of those in the absence of magnetic field. However, this is a function of the strength of magnetic field as well as the depth of the magnet system. In the following figure~\ref{fg2a}, the dependence on the magnetic field is shown for all depth values used in the analysis. A monotonic decrease in the relative fraction of cosmic rays leaking into the detector is observed as the field is increased. For a fixed value of field, higher depth corresponds to the higher degree of removal of cosmic rays. Similar trend is observed for electrons. This is shown in the adjacent plot~\ref{fg2b}.
\begin{figure}[ht]
\centering
\begin{subfigure}{.45\textwidth}
  \centering
  \captionsetup{justification=centering}
  \includegraphics[height=5.0 cm, width= 7.5 cm]{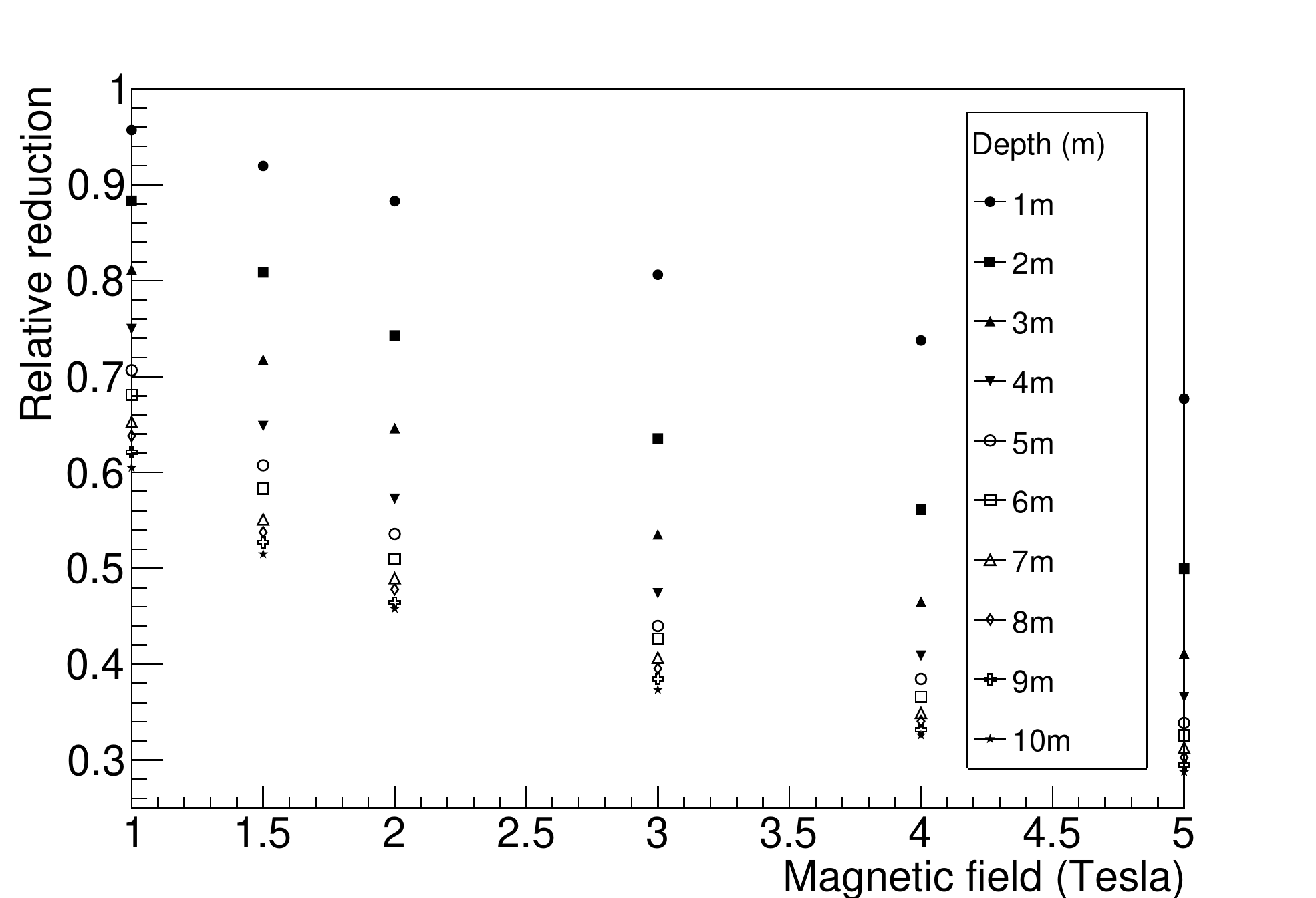}
  \caption{}
  \label{fg2a}
\end{subfigure}
\hspace{0.0 cm}
\begin{subfigure}{.45\textwidth}
  \centering
  \captionsetup{justification=centering}
  \includegraphics[height=5.0 cm, width= 7.5 cm]{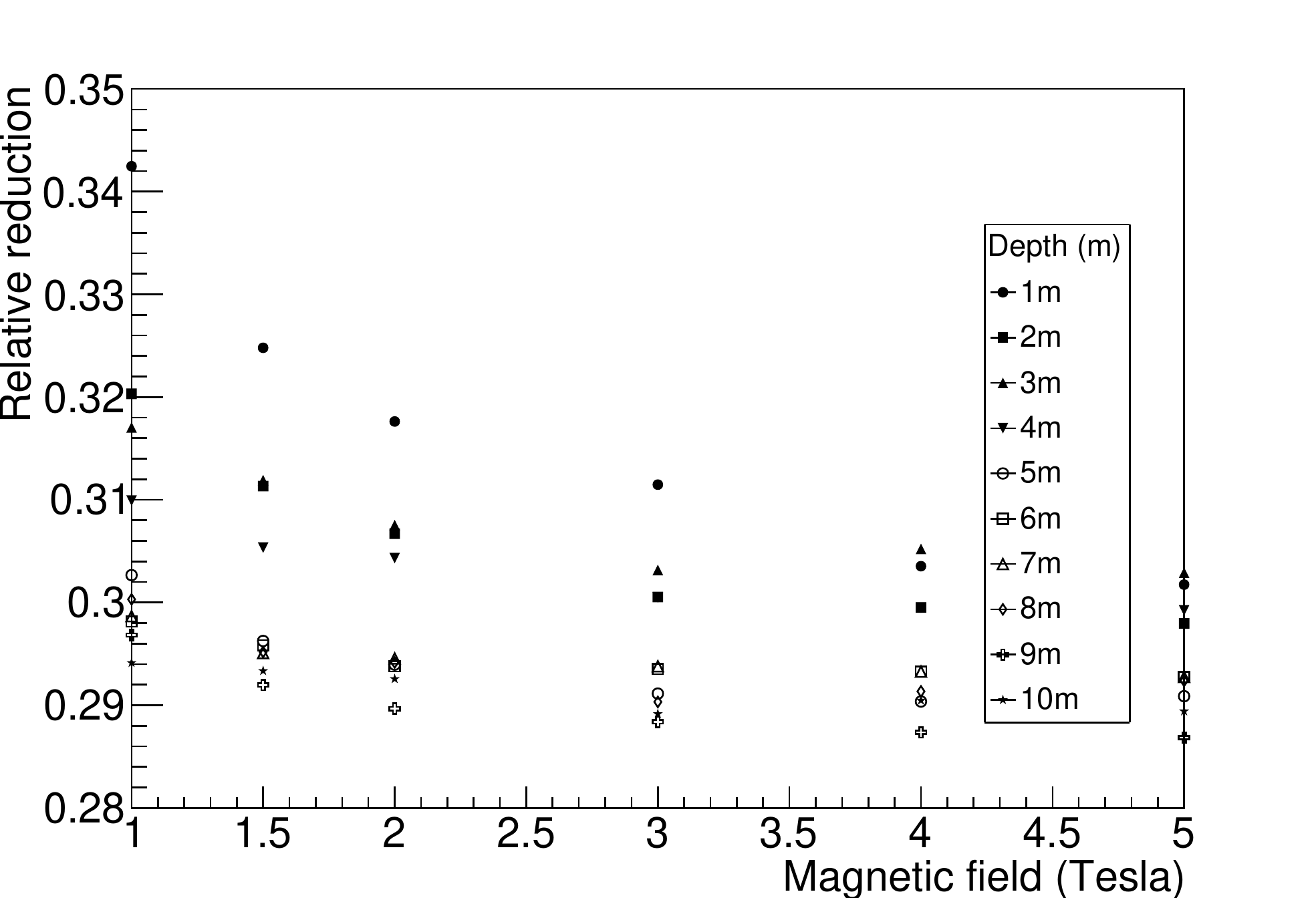}
  \caption{}
  \label{fg2b}
\end{subfigure}
\caption{(a) Effect of depth field strength on the relative reduction (ratio of number of muons which leaks into the detector in the presence and absence of magnetic field) of cosmic ray muons for various depth values (b) corresponding plot for electrons.}
\label{fig:2}
\end{figure}\\
Figure~\ref{fig:2} depicts the relative reduction in the total number of muons and electrons, without any reference to their energies. It does not reveal the energy bins which are more (or less) depleted when magnetic field is applied. However, this is an important parameter to consider while designing an experiment specifically when the signal region may be lying in in sub-GeV range. This is shown in the following figure~\ref{Spec_bin}. 
\begin{figure}[ht]
\centering
\begin{subfigure}{.45\textwidth}
  \centering
  \captionsetup{justification=centering}
  \includegraphics[height=5.0 cm, width= 7.5 cm]{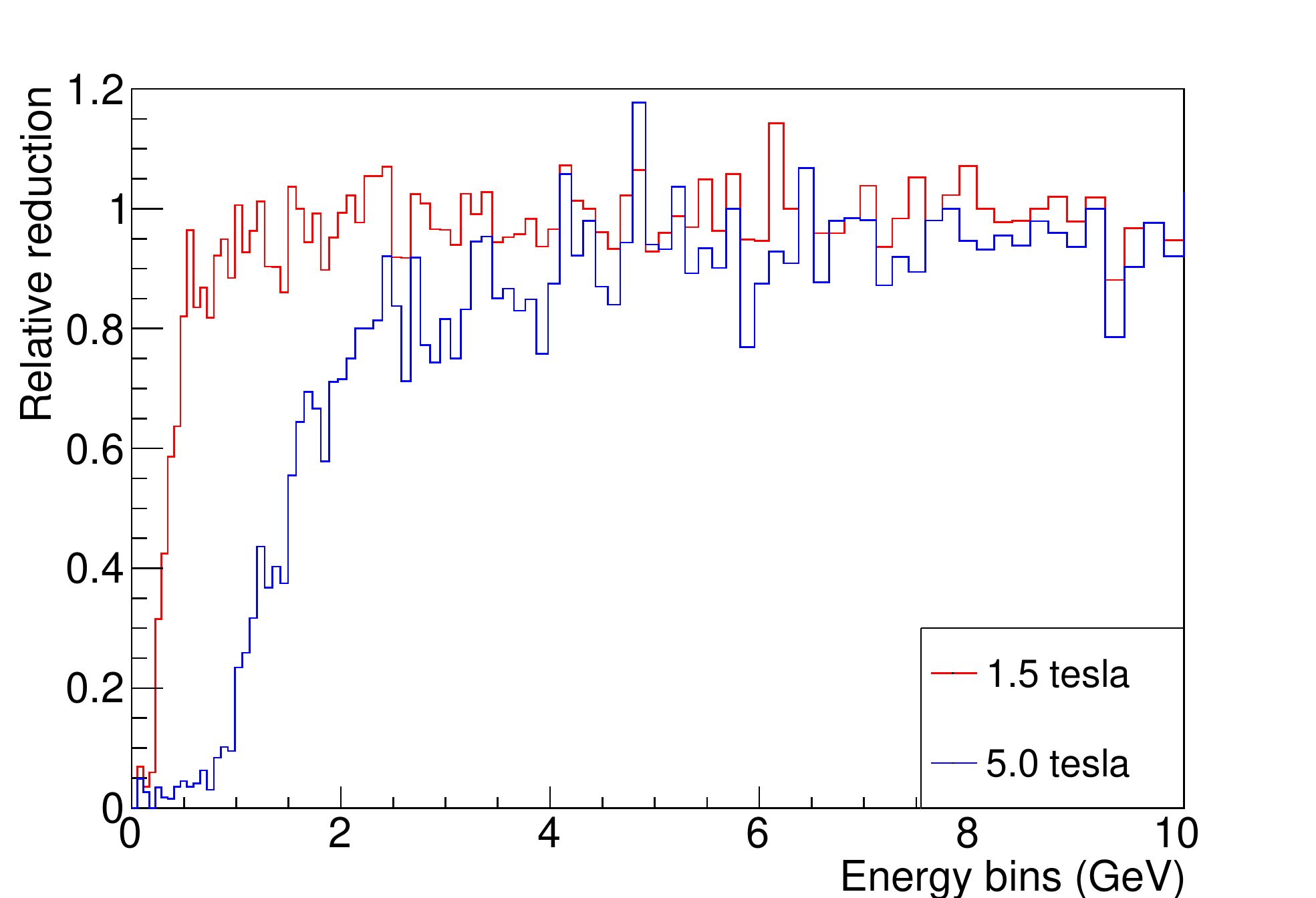}
  \caption{}
  \label{Spec_bin}
\end{subfigure}
\hspace{0.0 cm}
\begin{subfigure}{.45\textwidth}
  \centering
  \captionsetup{justification=centering}
  \includegraphics[height=5.0 cm, width= 7.5 cm]{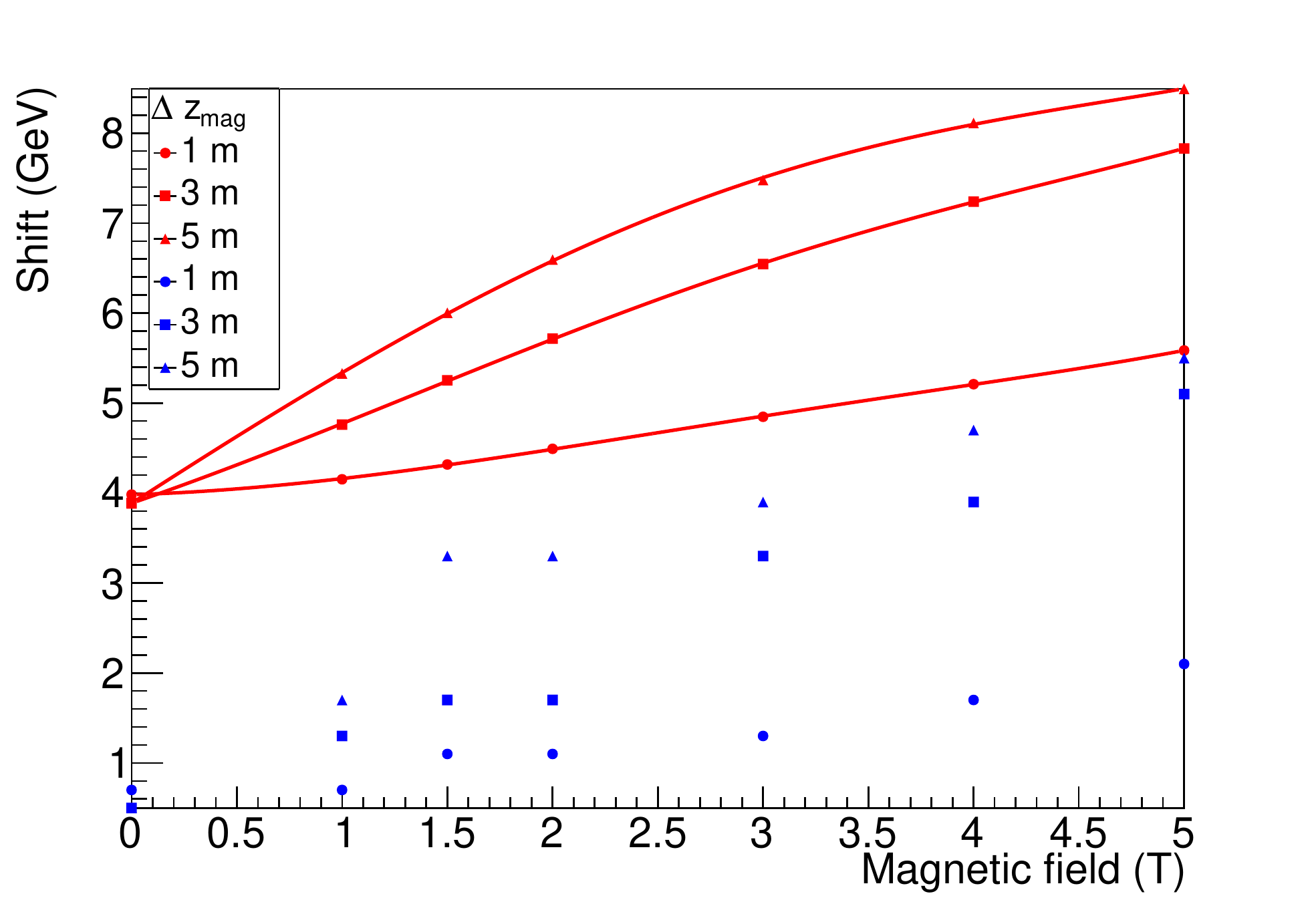}
  \caption{}
  \label{Shift}
\end{subfigure}
\caption{(a) Energy bin representation of relative reduction for two magnetic fields operating at 1 m depth; (b) systematic shift of mean of muon spectrum.}
\label{specEv}
\end{figure}

\subsection{Conclusion}
The preceding discussion quantifies the degree of 
reduction of cosmic ray charged particles possible through the use of magnetic field. As expected, a stronger field ensures less deep magnet with less volume. In a detector equipped with such a magnet overburden, if an observed event is reconstructed as a low energy event, it is much less probable to have come from cosmic rays. This conclusion is true whether or not the experiment is accelerator based.

The use of magnetic field shifts the residual cosmic muon spectrum to the higher energy end. From the perspective of the experimenter, the important parameters to consider are the mean and the mode (the energy bin where the events are populated the most) of the distribution. In the following figure~\ref{Shift}, the gradual increase of the mean (red) and the mode (blue) of the muon residual spectrum are presented by putting all the events below 20 GeV in 100 bins of equal width (so, each bin corresponds to 200 MeV). The increase in mean can be fitted with a polynomial of 5$^{th}$ order. 
The increase in mode with the increase in magnetic field is somewhat obscure due to the effect of the binning. But the overall trend is understandable.

\section{Transverse dimension (width) of magnet system}
In earlier study, the transverse dimension of the magnet system was taken as 10 m $\times$ 10 m. Does this parameter has a significant effect on the rate of charged particle cosmic background? To see this, GEANT4 simulation of 3 million events was performed for three different transverse dimensions of the magnet: (1) one whose width is the same as the particle detector, i.e. 5 m, (2) one whose width is 1.5 times that of the detector, i.e. 7.5 m and (3) one whose width is double, i.e. 10 m. The outcome of the experiment is shown in the following table~\ref{tab3}:
\begin{table}[H]
\begin{center}
\begin{tabular}{|l||*{3}{c|}}\hline
\backslashbox{depth}{width}
&\makebox[3em]{5 m}&\makebox[3em]{7.5 m}&\makebox[3em]{10 m} \\ \hline\hline
1 m & 11315 & 10493 & 10299\\\hline
2 m & 10455 & 9229 &  8893\\\hline
\end{tabular}
\caption{Effect of transverse dimension of the magnet on the residual flux of cosmic rays into the particle detector. The field strength is 1.5 tesla.}
\label{tab3}
\end{center}
\end{table}
So, increasing the width of the magnet system helps to reduce the rate of cosmic rays seen by the detector, as less number of particles can enter the detector from the side walls. However, the order of magnitude of the reduction is not very significant. Specifically, the reduction is about 1.8\% $\times$ depth (in meter) if the magnet transverse dimension is increased from 1.5 times to 2 times of the detector width.

\section{Magnet system}
It is known that many particle physics experiments are engineering strong magnets~\cite{ferracin2013development, ballarino2002current} etc. Can similar magnets be utilized to achieve this goal? Perhaps yes, but currently, these magnets do not enclose very large volumes and are not ideal for reducing the cosmic muon background. But the study presented in this paper gives a hint how the future magnets can be designed to achieve a desired degree of the reduction in residual cosmic muons. It is not only the strength that matters, the depth of the magnet is also a very important parameter.

The studies were presented mostly using a nominal field strength of 1.5 T which is attainable by an iron-core electromagnet or by rare-earth neodymium magnet. Use of iron or neodymium as the material of the magnet would lead to additional energy loss and multiple scattering of the charged particles of cosmic rays, apart from their desired deflection. This would make the role of magnetic field on the reduction in the cosmic rays obscure. This is why the preceding discussion assumed the field to be operating in vacuum. The effect of using the solid materials as magnet has been studied as well and is presented in the next subsection~\ref{fe}.

The superconducting magnets are scientifically the best options, as expected, since the field strength can be made very high (5-10 T, or even more). But with the current technology, the construction of a large volume superconducting magnet (even of a size comparable to the dimension of the detector) may be quite costly. But this cost must be compared with the total cost of construction and operation of the underground facilities and the physics output. This cost-benefit ratio will be discussed in section~\ref{costing}.

\subsection{Material of magnet system}\label{fe}
The cheapest option is perhaps to use a DC-based electromagnet with iron core. Another option is to use permanent neodymium bar magnets. One must keep in mind that the system must be supported on the surrounding ground due to their weight. Simulation of three million CRY cosmic ray events through 1 meter deep and 10 m$\times$ 10 m wide magnet system (for 1.5 T magnetic field) in 2.379 seconds produces a lot of secondary particles~\ref{tab4}:
\begin{table}[H]
\begin{center}
\begin{tabular}{|l||*{6}{c|}}\hline
\backslashbox{material}{particles}
 & $\mu$ & $e$ & $\gamma$ & $\nu_e$ & $\nu_\mu$ & n\\ \hline\hline
 Iron & 7240 & 1052 & 10677 & 1103 & 1860 & 599\\ \hline
 Neodymium & 7893 & 1218 & 12040 & 870 & 1519 & 1349\\ \hline
\end{tabular}
\caption{Effect of using solid ferromagnetic material to construct the magnet system.}
\label{tab4}
\end{center}
\end{table}
Comparison with table~\ref{tab2} shows that indeed the rate of muon flux decreases by $\sim27\%(20\%)$ for iron (Neodymium). However, that reduction gets compensated and surpassed by an increase in number of gamma rays, electrons, neutrinos and neutrons.
\begin{figure}[!ht]
\centering
\begin{subfigure}{.45\textwidth}
  \centering
  \captionsetup{justification=centering}
  \includegraphics[height=5.0 cm, width= 7.5 cm]{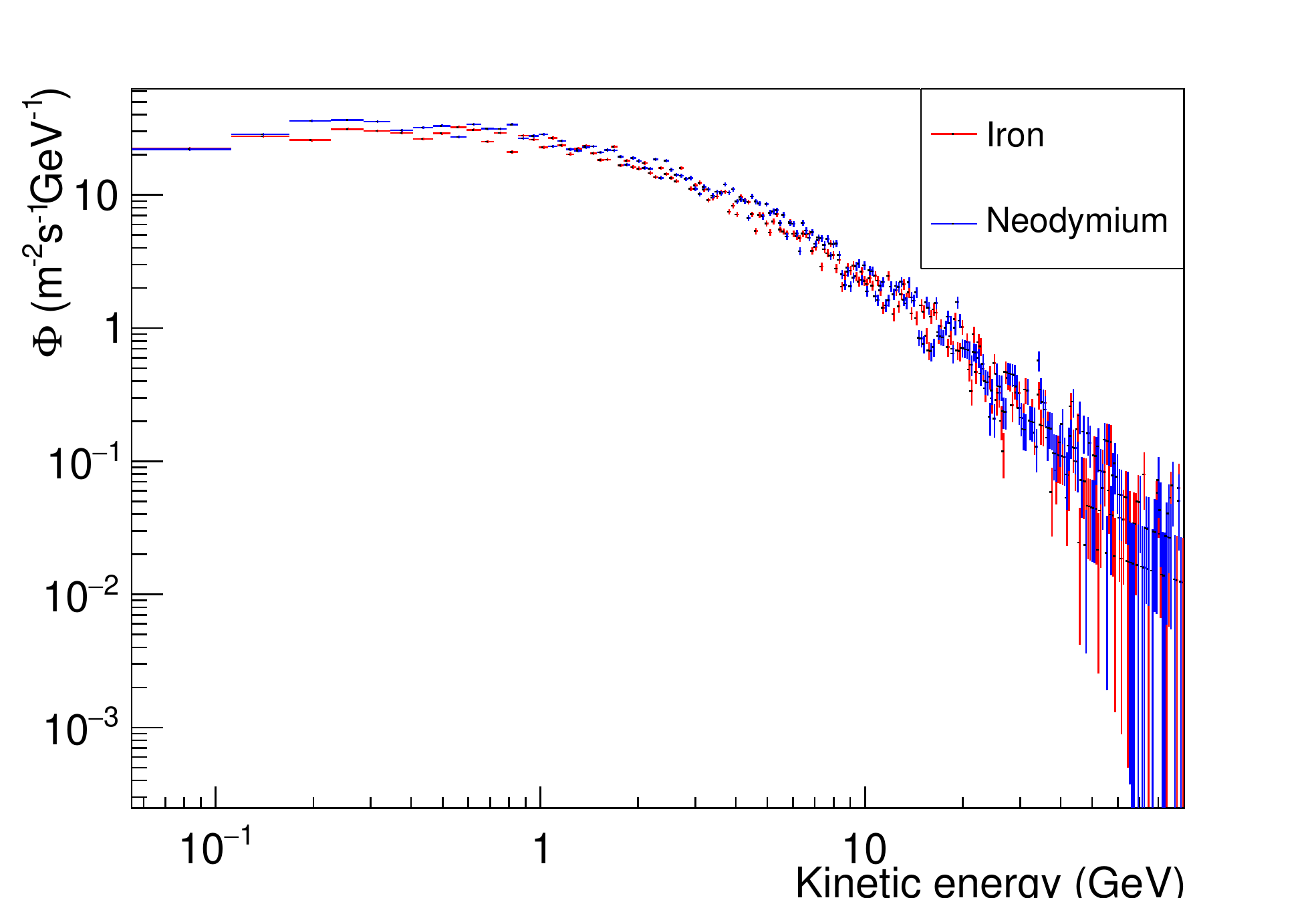}
  \caption{}
  \label{7a}
\end{subfigure}
\hspace{0.0 cm}
\begin{subfigure}{.45\textwidth}
  \centering
  \captionsetup{justification=centering}
  \includegraphics[height=5.0 cm, width= 7.5 cm]{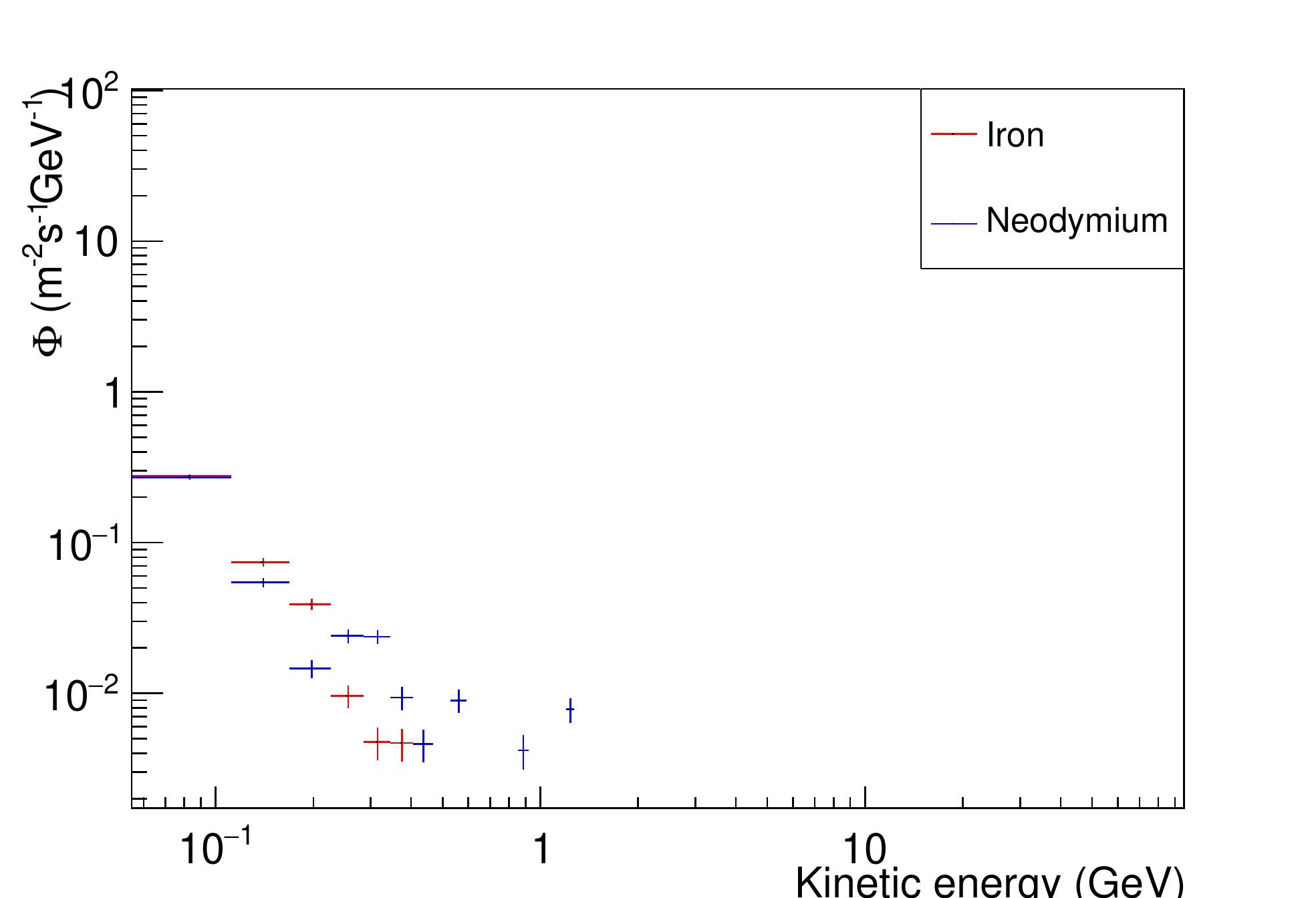}
  \caption{}
  \label{7b}
\end{subfigure}
\caption{Spectrum of (a) muons and (b) gamma which arise when cosmic rays are attempted to be blocked and deflected by 1 m deep iron/neodymium plates.}
\label{fig7}
\end{figure}
The following figure~\ref{fig7} shows the spectra of muons and photons arising out of the magnet which enter the detector. The former plot does not show any dip at low energy. Clearly, the effect of the energy loss in these solid materials dominates the effect of deflection due to magnetic field. The events in the lower energy bins had, on an average, an initial kinetic energy of ($\sim$1.6 MeV$\cdot$ cm$^2$/gm $\times$ 7.874 gm/cm$^3\times$ 100 cm$\sim$1.26 GeV) before entering the iron block. So, there is an overall down shift of spectrum and no specific dip is observed at low energy. The latter plot is added since many neutrino experiments detect neutrino signature by observing Cherenkov or scintillation light from neutrino events. The gamma rays coming from the magnetized block (placed above to deflect cosmic rays) would give rise to secondary backgrounds from gammas of energy up to 1 GeV. Not only that, the method also produces $\nu_\mu$ and $\nu_e$s which can also act as additional background to a low-medium energy neutrino experiment. Spectra of such neutrinos are shown in the following figure:~\ref{fig8}.
\begin{figure}[!ht]
\centering
\begin{subfigure}{.45\textwidth}
  \centering
  \captionsetup{justification=centering}
  \includegraphics[height=5.0 cm, width= 7.5 cm]{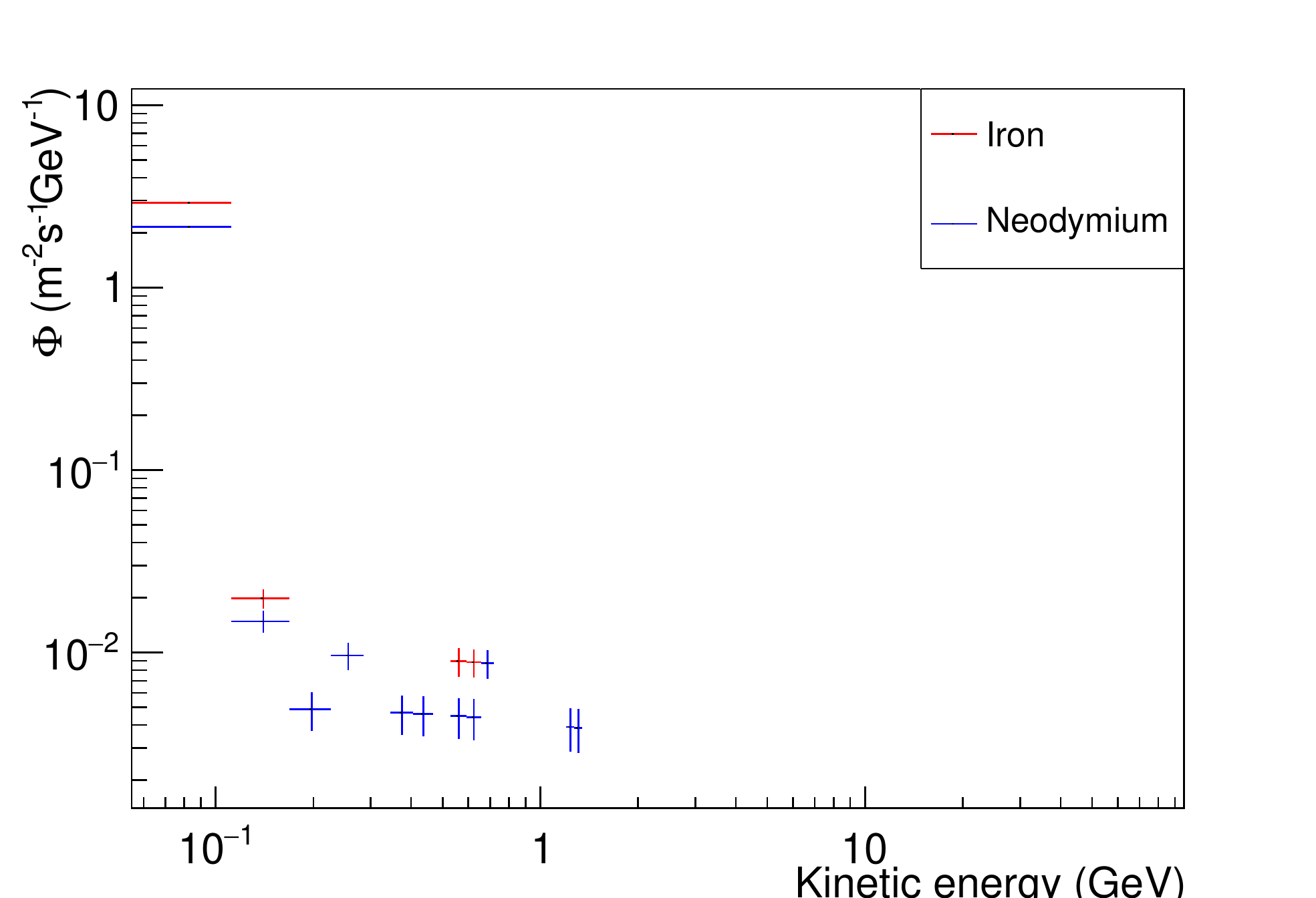}
  \caption{}
  \label{8a}
\end{subfigure}
\hspace{0.0 cm}
\begin{subfigure}{.45\textwidth}
  \centering
  \captionsetup{justification=centering}
  \includegraphics[height=5.0 cm, width= 7.5 cm]{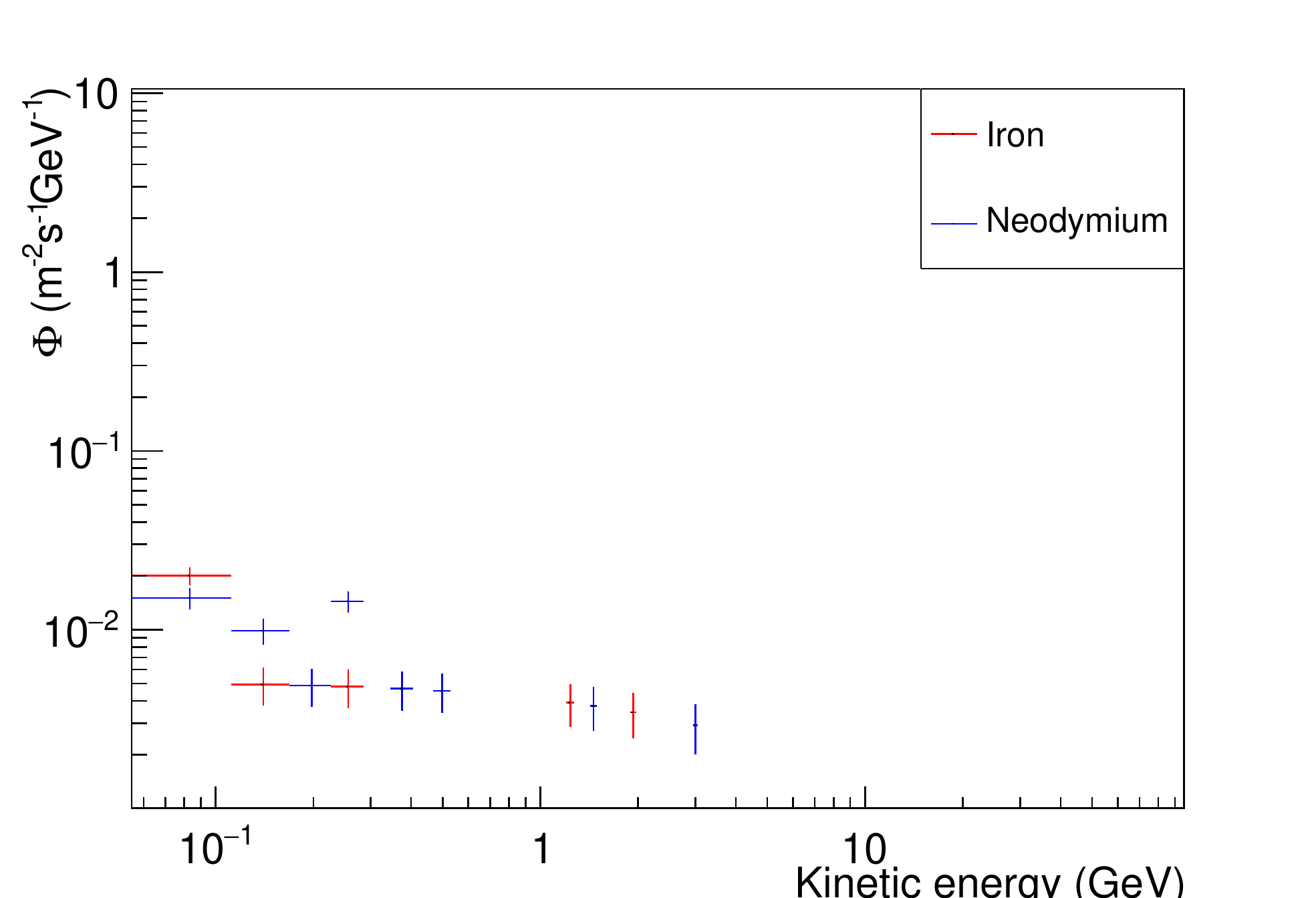}
  \caption{}
  \label{8b}
\end{subfigure}
\caption{Spectrum of (a) $\nu_\mu$ and (b) $\nu_e$ which arise when cosmic rays are attempted to be blocked and deflected by 1 m deep iron/neodymium plates.}
\label{fig8}
\end{figure}
The spike in the $\nu_\mu$ plot at $<100$ MeV bin is due to the decay of stopped pions which is monochromatic with 29.8 MeV energy. Although there is negligible number of pions in the cosmic rays at the ground level, the protons present in the cosmic ray produce large number of pions as they hit the block. The neutrinos with energy higher than $>100$ MeV are coincident with the lower side of standard atmospheric neutrinos.
\subsection{Conclusion}
A ferromagnetic material-based magnet will suppress the overall cosmic muon background, but not in an energy specific manner. It can be unambiguously stated, though, that a vacuum-core magnet can be used to reduce the cosmic muon and electron background significantly for a low energy ($<$100 MeV) nuclear or particle physics detector whose signals may come from particles with zero to tens of MeV of energy. The perfect examples of this kind of experiments are the reactor neutrino experiments and the current generation experiments (e.g. COHERENT) that intend to observe the ``coherent elastic neutrino nucleus scattering'' (CE$\nu$NS) and neutrinos coming from Supernovae (with tens of MeV of energy). The  COHERENT detector can achieve background rejection with timing of the neutrino pulse. But this is not possible while trying to observe neutrinos coming from Supernovae. The use of a magnet system can be fruitful to achieve very high degree of cosmic ray reduction. It may even present the possibility of performing event by event analysis at low energy bins.
\section{Comparison with standard rock overburden}
The discussion remains incomplete until we compare the preceding observation with the usual situation where a detector is constructed in an underground laboratory. Even a shallow underground laboratory receives much less number of muons~\cite{bogdanova2006cosmic}. However, along with the muons, other particles are also generated like the case of iron or neodymium. These include muon induced spallation neutrons which is a major issue for the direct dark matter detection experiments. The following table~\ref{tab5} shows the number of different particles as the depth of rock overburden of a shallow underground detector is increased.

\begin{table}[H]
\begin{center}
\begin{tabular}{|l||*{6}{c|}}\hline
\backslashbox{depth}{particles}
 & $\mu$ & $e$ & $\gamma$ & $\nu_e$ & $\nu_\mu$ & n\\ \hline\hline
 10 m & 5074 & 1086 & 6747 & 1045 & 1460 & 56\\ \hline
 20 m & 4620 & 1024 & 6709 & 1014 & 1265 & 73\\ \hline
 30 m & 4560 & 1004 & 6669 & 1040 & 1230 & 59\\ \hline
 40 m & 4239 & 874 &  5730 & 982 & 1246 & 39\\ \hline
 50 m & 3884 & 814 &  5727 & 981 & 1203 & 55\\ \hline
\end{tabular}
\caption{Number and composition of background cosmic rays to a 5 m $\times$ 5 m detector buried in underground in a span of 2.379 seconds (comprising 3 million CRY events).}
\label{tab5}
\end{center}
\end{table}
Comparison with table~\ref{tab4} shows that 10 m depth of rock overburden reduces muons by almost 30\% with respect to 1 m deep iron. The background of neutrons are also suppressed significantly. This shows why all the direct dark matter detectors are constructed underground. The spectrum of the muons as a function of shallow underground depth is shown in the following figure~\ref{9a}. 
\begin{figure}[!ht]
\centering
\begin{subfigure}{.45\textwidth}
  \centering
  \captionsetup{justification=centering}
  \includegraphics[height=5.0 cm, width= 7.5 cm]{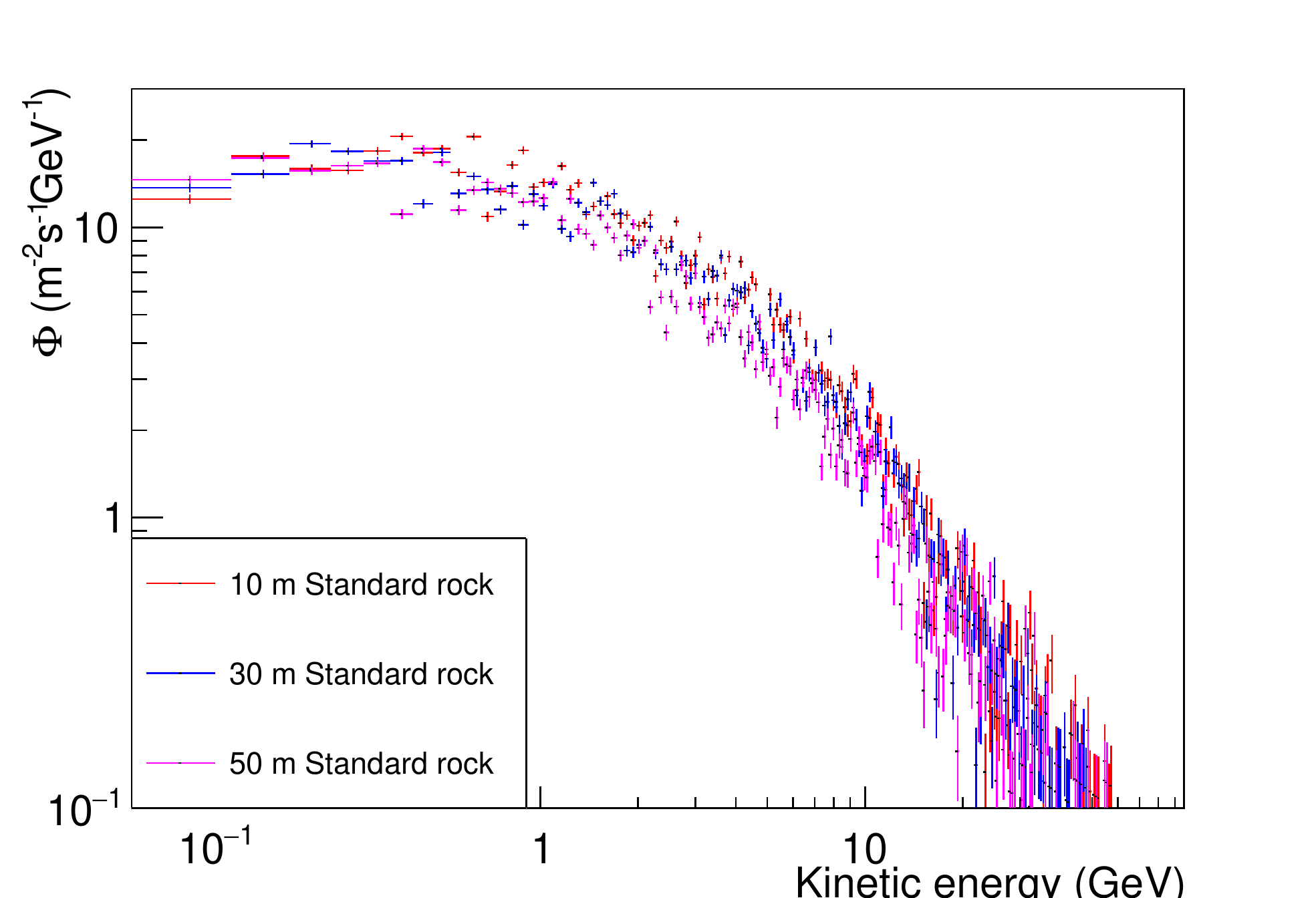}
  \caption{}
  \label{9a}
\end{subfigure}
\hspace{0.0 cm}
\begin{subfigure}{.45\textwidth}
  \centering
  \captionsetup{justification=centering}
  \includegraphics[height=5.0 cm, width= 7.5 cm]{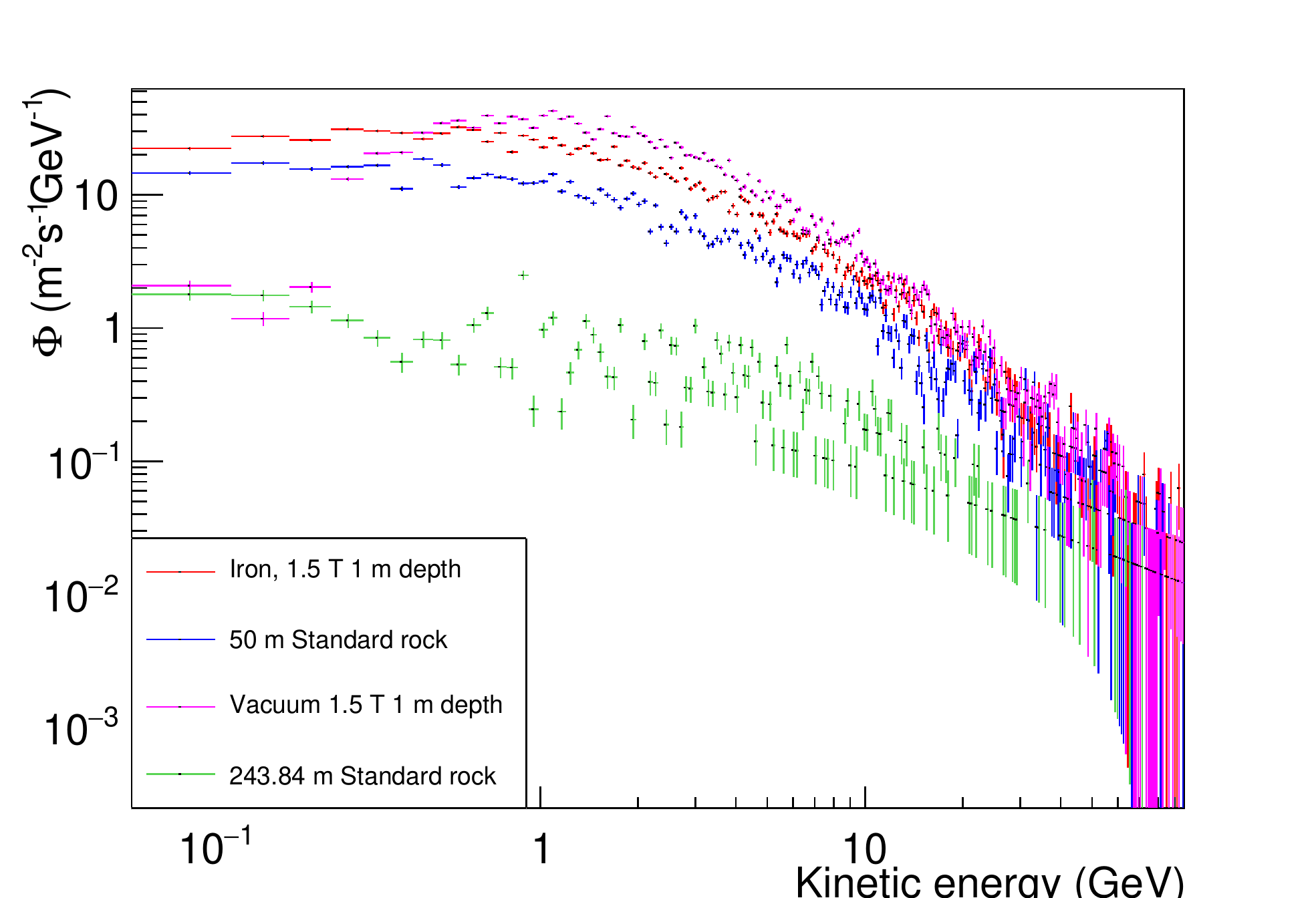}
  \caption{}
  \label{9b}
\end{subfigure}
\caption{(a) Spectrum of residual cosmic muons to an underground detector as a function of standard rock depth; rock density is taken as 2.65 g/cm$^3$. (b) comparison of the muon spectra for 1 m deep magnetic field in vacuum (purple), iron (red) with 50 m deep rock (blue) and 243.84 m (800 ft) deep rock (green).}
\label{fig8}
\end{figure}
The adjacent figure~\ref{9b} shows the comparison of the spectra of residual muons to a ground-based detector for 1.5 tesla field operating in 1 m deep magnet system in vacuum and in iron with 50 m and 800 feet deep (Homestake shallow level) underground detectors. The interesting point to see is the dip at low energy bins ($\sim $100 MeV) for the vacuum core magnet system. There is an overall suppression of cosmic muon background for iron/rock overburden (the deeper underground that laboratory is situated, the better). But there is no dip in these cases which reflects an overall down shift of the kinetic energy of all the cosmic rays. In comparison, it is seen that at $\sim 100$ MeV energy, the vacuum core magnetic field (purple) is almost as effective in reducing the cosmic muon background as a 800 ft. deep underground laboratory (green). In the case of the latter, the high energy tail is also suppressed as expected. If a stronger magnetic field can be constructed by deploying the superconducting magnet systems, it may be possible to reduce the cosmic muon background even further, extending to higher energy range.

\section{Cost comparison}\label{costing}
The next obvious point to consider is whether the use of a suitable  magnet system is cost-worthy in comparison with the construction of an underground detector. This comparison is somewhat difficult, because the available information on the cost of underground facilities reflect both the underground depths as well as the scales of the experiments. Not only that, the type of backgrounds received and their spectra at deep underground laboratories are very different from the composition and spectra of 
residual background in a detector placed under the magnet system. Nevertheless, a comparison may give some idea about the practicality of the use of the magnet systems. In the following table~\ref{tab6}, the range of costs of various underground research laboratories is shown (as found in~\cite{minecost} and~\cite{BureURL}).
\begin{table}[]
\begin{center}
\begin{tabular}{ccccc}\hline
Experiment & location & depth (m) & facility cost & annual operating cost\\ \hline
DM+DBD & Homestake & 2255 & 290-530 & 20\\ \hline
DM+DBD & SNOlab    & 2070 & 60 & n/a      \\ \hline
LBNE w/LAr-DM+DBD & Homestake & 1480 & 978-1137 & 18-23\\ \hline
\makecell{Study 155 million \\ year old clay rock} & Bure (France) & 450-500 & 315 & 67.65\\ \hline
\end{tabular}
\caption{Costs (2011 M\$) of construction and operation of the underground research laboratories at various depths. The acronyms DM, DBD and LAr in the above table correspond to dark matter, double $\beta$ decay and Liquid Argon. The cost at SNOlab is substantially less, presumably due to existing infrastructure. The cost for Bure URL has been converted from Euro to US\$.}
\label{tab6}
\end{center}
\end{table}
It would be nice to compare with the corresponding numbers for the shallow underground laboratories (e.g. Felenskeller laboratory at 47 m depth and the shallow underground laboratory at Pacific Northwest National Laboratory at 11.3 m depth). But numbers representing the cost were not found from reliable resources.\\
The cost of superconducting magnet as a function of the field volume times the field strength has been discussed in the literature~\cite{green2008cost}. From the figures and the formulae given in these papers, we see that the cost of the superconducting magnet of volume-strength 150 Tm$^3$ will be about 15-25 M\$ in accordance with 2008 valuation. This is an order of magnitude less than the costs of the major underground laboratories.
\section{Indirect implication of the method for WIMP dark matter detectors}
The detectors for the direct search of dark matter 
are usually placed underground for reducing cosmic ray background to as minimum as possible. They look for Weakly Interacting Massive Particles (WIMPs) that are hypothesized to be electrically neutral particles only responding to weak interaction apart from gravity. For such detectors, perhaps neutrons are the most problematic backgrounds, as they lead to nuclear recoil events exactly the same way as could be done by the WIMPs. A magnet system cannot deflect away electrically neutral neutrons of the cosmic rays. So, it is not directly helpful for such  detectors. However, one of the sources of background neutrons in the underground dark matter detectors is the spallation of cosmic muons and the other charged particles in the earth's crust. This continues to be a background, even if the detector is buried deep in the underground. Multiple levels of veto (e.g. water shielding) and reconstruction are needed to take care of the neutron background. A reduction in the residual cosmic muon rays at the ground level which can be achieved by the magnet system, can partially reduce the neutron background due to the spallation neutrons to an underground dark matter detector. This is because, a less fraction of muons will lead to a less fraction of spallation neutrons. Detectors that intends to observe solar neutrinos or diffuse Supernova neutrino background (DSNB) signals~\cite{li2014first} may also benefit from this technique. In this case, it may be helpful to bury the detector deep in the underground with the magnet resting on top of the ground.

\section{Summary}
In this work, it has been shown that a ground based magnet system may be scientifically useful to cut down the background due to the cosmic ray charged particles. With currently available technology, it is possible to reduce the cosmic background to low energy neutrino detectors. A significant reduction of the higher energy cosmic muons will require a stronger superconducting magnet. The technique can also indirectly help in reducing the background of spallation neutrons at deep underground detectors. It will be a sheer engineering challenge to build a large superconducting magnet and to make sure that the field does not leak, but it will save the cost of constructing an underground detector and danger associated with it.

\section{Acknowledgement}
Manipal Centre for Natural Sciences, Manipal Academy of Higher Education,Manipal is acknowledged for the encouragement and facilities to carry out this work.
\bibliographystyle{unsrt}
\bibliography{CosmicRayBkgReduction}
\end{document}